\documentstyle[prd,aps]{revtex}
        		
\def\lsim{\mathrel{\rlap{\lower4pt\hbox{\hskip1pt$\sim$}}
    \raise1pt\hbox{$<$}}}
\def\gsim{\mathrel{\rlap{\lower4pt\hbox{\hskip1pt$\sim$}}
    \raise1pt\hbox{$>$}}}
\def\mpl{M_{{\rm Pl}}}

\newcommand\eq[1]{Eq.~(\ref{#1})}
\newcommand\eqs[2]{Eqs.~(\ref{#1}) and (\ref{#2})}

\newcommand\pa{\partial}

\def\dlabel#1{\label{#1} \ \ \ \ \ \ \ \ #1}

\newcommand\sub[1]{_{\rm #1}}

\newcommand\ee{\end{equation}}
\newcommand\be{\begin{equation}}
\newcommand\eea{\end{eqnarray}}
\newcommand\bea{\begin{eqnarray}}

\newcommand\GeV{\,\rm{GeV}}

\newcommand\Mpl{M_{{\rm Pl}}}

\def\dslash{\not{\hbox{\kern-2pt $\partial$}}}
\def\Dslash{\not{\hbox{\kern-4pt $D$}}}
\def\Oslash{\not{\hbox{\kern-4pt $O$}}}
\def\Qslash{\not{\hbox{\kern-4pt $Q$}}}
\def\pslash{\not{\hbox{\kern-2.3pt $p$}}}
\def\kslash{\not{\hbox{\kern-2.3pt $k$}}}
\def\qslash{\not{\hbox{\kern-2.3pt $q$}}}
 \newtoks\slashfraction
 \slashfraction={.13}
 \def\slash#1{\setbox0\hbox{$ #1 $}
 \setbox0\hbox to \the\slashfraction\wd0{\hss \box0}/\box0 }
 
\def\eeq{\end{equation}}
\def\beq{\begin{equation}}

\begin{document}

\preprint{LANCS-TH/9716,hep-ph/9710347}
\draft
\tightenlines

\title{Non-renormalizable terms and M theory during inflation}
\author{David H.  Lyth}
\address{Department of Physics, \\
University of Lancaster, Lancaster LA1 4YB U.~K.}

\date{October 1997}
\maketitle

\begin{abstract}
Inflation is well known to be difficult in
the context of supergravity, 
if the potential is dominated by the
$F$ term. Non-renormalizable terms
generically give $|V''|\sim V/M^2$, where
$V(\phi)$ is the inflaton potential and
$M$ is the scale above which the effective field theory 
under consideration is supposed to break down.
This is equivalent to $|\eta|\sim (\mpl/M)^2 > 1$ where
$\mpl=(8\pi G)^{-1/2}$, but inflation requires $|\eta|<0.1 $. 
I here point out that all of the above applies also if the $D$ term
dominates, with
the crucial difference that the generic result is now
easily avoided by imposing a discrete symmetry.
I also point out that if extra spacetime dimensions appear well below
the Planck scale, as in a 
recent M-theory model, one expects $M\ll \mpl$,
which makes the problem worse than if $M\sim \mpl$.
\end{abstract}
\newpage
\baselineskip=20pt

{\bf 1.}~~ To achieve slow-roll inflation, the potential $V(\phi)$
must satisfy the flatness conditions $\epsilon\ll 1$
and $|\eta|\ll 1$, where \cite{LL2}
\bea
\epsilon &\equiv & \frac12 \mpl^2 (V'/V)^2 \\
\eta &\equiv & \mpl^2 V''/V
\eea
and $\mpl = (8\pi G)^{-1/2} = 2.4\times 10^{18}\GeV$ is the reduced
Planck mass.  When these are satisfied, the time dependence of the
inflaton $\phi$ is generally given by the slow-roll expression
$3H\dot\phi = -V'$, where $H\simeq \sqrt{\frac13\Mpl^{-2} V}$ 
is the Hubble parameter during inflation.
On a given scale, the spectrum of the primordial curvature perturbation,
thought to be the origin of structure in the Universe,
is given by
\be
\delta_H^2 (k) = \frac1{150\pi^2 \mpl^4} \frac V \epsilon
\ee
The right hand side is evaluated when the relevant scale
$k$ leaves the horizon. On large scales, the COBE observation
of the cmb anisotropy corresponds to
\be
V^{1/4}/\epsilon^{1/4} = .027\mpl = 6.7 \times 10^{16}\GeV
\label{cobe}
\ee

The spectral index of the primordial curvature perturbation
is given by
\be
n-1 = 2\eta - 6\epsilon 
\ee
A perfectly scale-independent spectrum would correspond to $n=1$, 
and observation already demands $|n-1| < 0.2$. Thus $\epsilon$ 
and $\eta$ have to be $\lsim 0.1$ (barring a cancellation)
and this constraint will get tighter
if future observations move $n$ closer to 1. Many models of inflation
predict that this will be the case, some giving a value of $n$ 
completely indistinguishable from 1.

Usually, $\phi$ is supposed to be charged under at least a
$Z_2$ symmetry $\phi\to-\phi$, which is unbroken during inflation.
Then $V'=0$ at the origin, and inflation typically takes place near the 
origin. As a result $\epsilon$ negligible compared with $\eta$, and
$n-1 = 2\eta\equiv 2\mpl^2 V''/V$. We assume that this is the case in 
what follows. If it is not, the
nonrenormalizable terms generically give both $|\eta|\sim 1$
{\em and} $\epsilon\sim 1$ at a generic point in field space,
making model-building even more tricky. 

{\bf 2}~~ In supergravity, the tree-level potential
is the sum of an $F$-term and a $D$-term \cite{susy}, 
\be
V=V_F +V_D
\ee
Supergravity is a non-renormalizable field theory, whose
lagrangian presumably contains an infinite number of non-renormalizable
 terms.
Taking the theory to be an effective one, holding up to some scale $M$,
the coefficients of the non-renormalizable terms are expected to be
generically of 
order 1, in units of $M$. Hopefully, they can be calculated if one 
understands what goes on at scales above $M$.
The most optimistic view is to take $M$ to be the scale above
which field theory itself breaks down, due to gravitational effects such 
as the quantum fluctuation
in the spacetime metric.
It is not unreasonable to take this view in the 
context of inflation, if the inflaton is a gauge
singlet, and we adopt it here.\footnote
{For fields charged under the Standard Model gauge interactions
one should use $M=M\sub{GUT}\simeq 10^{16}\GeV$ in the low energy theory,
and for hidden/secluded sector fields below the condensation scale
$\Lambda$ one should use $M=\Lambda$.}
With four spacetime dimensions, this means that 
$M=\mpl$. If more dimensions open up well below the Planck scale, 
the answer is less obvious (see below).

The non-renormalizable terms respect the gauge symmetries
possessed by the renormalizable theory, but they need not
respect its global symmetries. On the contrary, it 
is generally felt that the usual continuous global
symmetries  (built out of $U(1)$'s acting on the phases of the complex 
fields) will {\em not} generically be respected, at least if 
$M$ is the scale at which gravitational effects spoil field theory.
This viewpoint derives in part from superstring theory, but
that theory also suggests that discrete subgroups of the global symmetries
(built out of $Z_N$'s)
{\em will} occur. By imposing suitable discrete symmetries, one can ensure that
a global continuous symmetry is approximately preserved at field
values $\ll M$ \cite{dine}.
This is important,
because several cases are known where 
an approximate global $U(1)$ is phenomenological desirable.
The best known case is Peccei-Quinn symmetry, which must be preserved
to high accuracy in order to keep the axion sufficiently light.

Some time ago \cite{CLLSW}, it was emphasized 
that non-renormalizable terms will contribute to $V_F$, 
generically giving
$|V_F'' | \sim V_F/M^2$.\footnote
{The result holds actually for the second derivative in any field
direction, as was 
noted a long time ago \cite{dinefisch,coughlan}.}
Most models of inflation have the $F$-term dominating
($F$-term inflation)
and then the non-renormalizable terms in $V_F$ give
\be
|\eta|\sim \mpl^2/M^2 
\ee
This generic result is too big, since
slow-roll inflation {\em per se} 
requires $|\eta|\ll 1$, and the observational bound on 
$n$ requires $|\eta| <0.1$

A little later \cite{ewansgrav} 
it was pointed out that the problem may disappear 
if instead $V_D$ dominates ($D$-term inflation).
It was also pointed out that $D$-term {\em hybrid} inflation occurs quite 
naturally, if the lagrangian contains a Fayet-Illiopoulos term.
(Within the single-field paradigm, $D$-term inflation 
is essentially impossible, because the inflaton then has to be charged under 
the relevant $U(1)$ and the gauge couplings spoil inflation unless
they are unreasonably small \cite{casas}.)
Several authors 
\cite{bindvali,halyo,j2,dvaliriotto,casas2,lr,steve}
have since studied models of $D$-term inflation, always
under the same assumption that non-renormalizable terms can be ignored
in that case.

Here I point out that non-renormalizable terms also contribute to
$V_D$, through the gauge kinetic function.
They {\em generically} give $|\eta|\sim (\mpl/M)^2$ in that case also,
but in contrast with the case of the $F$ term this generic result
is easy to evade by imposing a discrete symmetry. The reason is that
the gauge kinetic function appearing in the $D$ term 
is holomorphic, in contrast with the
Kahler potential which appears in the $F$ term.

I consider the usual model of $D$-term hybrid inflation
\cite{ewansgrav,bindvali,halyo,j2,dvaliriotto,lr,steve}.
There is just one non-vanishing $D$ field, which contains a 
Fayet-Illiopoulos term with coefficient $\xi$. The 
fields charged under the relevant local $U(1)$ have been driven 
to zero (or anyhow sufficiently small values)
along with $V_F$  Then
\be
V\simeq V_D \simeq \frac{\xi^2 g^2}{2} \,{\rm Re}\, f^{-1} 
\ee
where $g$ is the gauge coupling of the relevant $U(1)$, and the 
gauge kinetic function $f$ is a holomorphic function of all 
of the complex scalar fields  $\phi_n$.
If we regard $g^2$ as fixed, 
$f$ must be invariant under all internal symmetries. 

At a given point in field space,
one can choose $f=1$ corresponding to
canonical normalization of the gauge field of the relevant $U(1)$.
This point is conveniently taken to be the origin, defined as the fixed
point of the usual internal symmetries.
Then, along say the $\phi_1$ direction,
\be
1/f= 1 + \lambda
M^{-2} \phi_1^2 + \cdots
\ee
There is no linear term, unless $\phi_1$ is a singlet under all of the 
internal symmetries that are unbroken during inflation. 
The quadratic term is allowed if
the only symmetry is $\phi_1\to -\phi_1$.
Then, if inflaton is $\phi={\sqrt 2 \rm Re}\, \phi_1$,
it gives a contribution $\eta = \lambda (\mpl/M)^2$, with $\lambda$
generically of order 1.

The offending term is forbidden if there is
a $Z_N$ symmetry ($\phi_1\to \exp(i\alpha)\phi_1$ 
with $\alpha = 2\pi/N$) with $N\geq 3$. It is also forbidden if there
is a global
$U(1)$ symmetry, corresponding to arbitrary $\alpha$.
In all of the models of $D$-term inflation proposed so far,
such a global $U(1)$ is present as an $R$ symmetry,
with $W\propto \phi_1$. In the simplest case, 
\be
W=c \phi_1\phi_2\phi_3
\ee
where $\phi_2$ and $\phi_3$ oppositely charged 
under the relevant gauge $U(1)$. 
One indeed needs to forbid terms in $W$ of the form
$\phi_1^2\phi_n$ ($n\neq 1$)
because they would generate a quartic term in the inflaton 
potential and spoil inflation. But to achieve this it is enough
to have the symmetry $\phi_1\to -\phi_1$ (acting on $W$ as an $R$ symmetry).
I am here pointing out that this symmetry is {\em not} enough 
to forbid the disasterous quadratic term in $f$; it has to be promoted 
to $Z_{2N}$ with $N>1$, or to the full $U(1)$. One hopes that 
the superstring will ultimately determine which discrete symmetry actually
holds, if any.

In one model of $D$-term inflation \cite{steve}, the global $U(1)$ is the 
Peccei-Quinn symmetry, whose non-perturbative breaking
ensures the CP invariance
of the strong interaction. It has to be preserved to high accuracy
in order to keep the axion sufficiently light, and it appears
that discrete symmetries of the model ensure this.\footnote
{In an earlier version of this paper I stated that they
will not do so.}
Of course, this means that the quadratic term in $f$ is killed.
This model has the virtue of making
contact with both Peccei-Quinn symmetry and the Standard Model.
On the negative side, it so far lacks a definite origin for
the Fayet-Illiopoulos term; in contrast with the other models
of $D$-term inflation,
a direct origin in the superstring seems impossible since
$V^{1/4}$ is extremely low.

{\bf 3}~~ Let us briefly recall the situation for the $F$ term 
\cite{CLLSW}.
At least at tree level, 
\be
V_F = e^{K/\mpl^2} \left[
 \sum_{nm} \left(W_n+\Mpl^{-2}WK_n \right)K^{n \bar m} 
\left(\bar W_{\bar m} + \Mpl^{-2}\bar W K_{\bar m} \right)
-3\mpl^{-2}|W|^2\right]
\dlabel{vf}
\ee
Here, $W$ is the superpotential (holomorphic in the fields)
and $K$ is the Kahler potential (a non-singular 
function of the fields and their
complex conjugates). A subscript $n$ denotes
$\pa/\pa\phi_n$, and a subscript $\bar m$ denotes $\pa/\pa\bar\phi_m$.
Also, $K^{n\bar m}$ is
the matrix inverse of $K_{n\bar m}$.
Only the combination $G\equiv K + \ln|W|^2$ is physically significant.

Canonically normalizing the fields at (say) the origin corresponds to
\be
K = \sum _n|\phi_n|^2 + O(\phi_n^3)
\ee
(Any linear term can be absorbed into $W$.) Then,
\be
K^{n\bar m} = \delta_{nm} + M^{-2}
\sum_n \lambda_n |\phi_n|^2 + \cdots 
\ee
In the last expression, the 
terms not displayed are linear and higher; the quadratic term
displayed is a particularly simple one, which cannot be forbidden
by any of the usual symmetries.

Now make again the assumption that
$\phi={\rm Re}\,\phi_1/\sqrt2$; as we remark in a moment, the opposite
assumption that $\phi$ corresponds instead to the phase of $\phi_1$
would give something dramatically different. For simplicity, also
set
$e^K=1$ corresponding to small field values. Then, 
one can identify some contributions
to $V''$,
\be
V'' = \mpl^{-2} \left[ V  - |W_1|^2 \right] 
+ M^{-2} \left[ \sum_n \lambda_n |W_n|^2
+\cdots \right]
\label{sep}
\ee
In the first bracket, $V$ comes from differentiating $e^K$, and the
coefficient of $-|W_1|^2$ is the sum of $+2$ coming
from the first term in the bracket
of \eq{vf}, and $-3$ coming from the 
$-3\mpl^{-2}|W|^2$.
The three terms displayed will give a contribution
\be
\eta = 1 - a + b (\mpl/M)^2 
\label{eta}
\ee
with generically $|b|\sim 1$.

It was pointed out in \cite{CLLSW} if $W=\Lambda^2\phi_1$ 
during inflation, then
$a=1$. (It is easy to construct models of inflation where this is 
exactly \cite{CLLSW,dss} or approximately \cite{izawa} true.)
In that case, one need only require $|b|\ll 1$, corresponding
to an accidental suppression of the other non-renormalizable contributions. 
Some authors \cite{izawa,linderiotto,ddr}
have taken the view that this is an improvement
on the generic situation, where one needs in addition the
accident $|a-1|\ll 1$.

Notice that \eq{vf} contains $\mpl$, as distinct from the scale $M$
above which field theory is supposed to break down.
Taking $\mpl$ to infinity with $M$ fixed
converts supergravity into a non-renormalizable globally supersymmetric
theory. More usually, one considers the limit where $\mpl$ and $M$ go to 
infinity together, corresponding to a renormalizable globally
supersymmetric theory. It is clear from the form of \eqs{cobe}{eta}
that, during inflation,
neither of these prescriptions should be used without justification.
As we have just seen though, the use of 
non-renormalizable global
supersymmetry can be justified if the superpotential is linear.

If $b\sim 1$ (the generic case) the only way of evading the
result $|\eta|\sim (\mpl/M)^2$
is to suppose that $\phi$ is the pseudo-Goldstone boson of a global 
symmetry \cite{natural} acting on (say) $\phi_1$. 
Then all $|\phi_n|$ are fixed during inflation, and
if the symmetry is broken only by $W$ 
and not by $K$, the non-renormalizable terms in the latter
need cause no problem \cite{CLLSW}, though one still has the problem of 
understanding why a global continuous symmetry is respected by 
non-renormalizable terms.
A more promising avenue is to suppose that 
$K$ and $W$ have very special forms
\cite{ewansgrav,gmo,glm}, corresponding roughly to versions of
no-scale supergravity. In this case, it may be justified to use
the limit of renormalizable global supersymmetry.

{\bf 4.}~~ Finally, I note that the scale $M$ might not be as big
as $\mpl$ if additional space dimensions open up below
$\mpl$. I have in mind a specific example, where the underlying theory
is an M theory \cite{mtheory}. A field theory is valid below some 
scale $Q<\mpl$, and it is attractive to choose 
$Q\sim 10^{-2}\mpl\sim 2\times 10^{16}\GeV$ 
to account for the 
apparent unification of the gauge couplings. This field theory
lives in effectively five spacetime 
dimensions, until we get down to some still lower scale.
If the fifth dimension makes no significant 
difference, the scale $M$ of the non-renormalizable terms
will be $M=Q\sim 10^{-2}\mpl$. Otherwise $M$
will presumably be lower, though with two mass scales in the 
underlying theory the concept of a single scale $M$ may not even be 
useful. As yet nothing has been worked regarding these questions, even in 
the specific example mentioned. 

If $M$ is indeed of order $10^{16}\GeV$, a field-theory model of 
inflation will not make sense \cite{banksdine} if 
$V^{1/4}\sim 10^{16.5}\GeV$, the maximum allowed by the COBE
normalization \eq{cobe}. A {\em four-dimensional} field theory model
will not make sense unless $V^{1/4}$ is even lower.
However, many models have been proposed
with low $V^{1/4}$.

Of these, the simplest is the $D$-term model, with the slope of the
potential coming from the 1-loop correction
\cite{bindvali,halyo}. In that case 
COBE normalization requires \cite{j2,dvaliriotto}
\be
\frac{V^{1/4}}{(g^2/2)^{1/4}} = \sqrt \xi =
3\times 10^{15}\GeV
\ee
This might be low enough to justify the use of four-dimensional
field theory. Also, if
$\xi$ comes the superstring, it will be a loop suppression factor
times $M^2$, which might be compatible with the M theory model just 
mentioned. Again, this has yet to be investigated.

The most promising 
$F$-term model is probably that of \cite{ewanloop2}.
Starting at the scale $\phi = M$ with the generic 
$V''(\phi)\simeq (\mpl/M)^2 V$, renormalization group equations are invoked
to run this quantity to zero at a scale $\phi\ll M$, where inflation 
occurs. In this case one can have $V^{1/4}$ as low as $\sqrt{M m_s}$ where
$m_s=100\GeV$, amply justifying the use of four-dimensional field
 theory. 
As given, this model takes $M = \mpl$, and it 
is unclear whether the change 
$M\sim \mpl/100$ would make an important difference.

\vskip 1cm
\underline{Acknowledgements}:

I thank Toni Riotto for useful comments on a first draft of this paper.
This work is partially supported by grants from PPARC and
from the European Commission
under the Human Capital and Mobility programme, contract
No.~CHRX-CT94-0423. 

\def\NPB#1#2#3{Nucl. Phys. {\bf B#1}, #3 (19#2)}
\def\PLB#1#2#3{Phys. Lett. {\bf B#1}, #3 (19#2) }
\def\PLBold#1#2#3{Phys. Lett. {\bf#1B} (19#2) #3}
\def\PRD#1#2#3{Phys. Rev. {\bf D#1}, #3 (19#2) }
\def\PRL#1#2#3{Phys. Rev. Lett. {\bf#1} (19#2) #3}
\def\PRT#1#2#3{Phys. Rep. {\bf#1} (19#2) #3}
\def\ARAA#1#2#3{Ann. Rev. Astron. Astrophys. {\bf#1} (19#2) #3}
\def\ARNP#1#2#3{Ann. Rev. Nucl. Part. Sci. {\bf#1} (19#2) #3}
\def\MPL#1#2#3{Mod. Phys. Lett. {\bf #1} (19#2) #3}
\def\ZPC#1#2#3{Zeit. f\"ur Physik {\bf C#1} (19#2) #3}
\def\APJ#1#2#3{Ap. J. {\bf #1} (19#2) #3}
\def\AP#1#2#3{{Ann. Phys. } {\bf #1} (19#2) #3}
\def\RMP#1#2#3{{Rev. Mod. Phys. } {\bf #1} (19#2) #3}
\def\CMP#1#2#3{{Comm. Math. Phys. } {\bf #1} (19#2) #3}


\begin{thebibliography}{99}
\bibitem{LL2} For a review of this material concerning inflation, see
A. R. Liddle and D. H. Lyth, Phys. Rep. {\bf 231},
1 (1993).
\bibitem{susy} For reviews of supersymmetry and supergravity, see
        H. P. Nilles, Phys. Rep. {\bf 110}, 1 (1984); H.E. Haber and
G.L. Kane, Phys. Rept. {\bf 117}, 75 (1985) and
        D. Bailin and A. Love, {\em Supersymmetric Gauge Field Theory
        and String Theory}, IOP, Bristol (1994).
\bibitem{dine} A useful review of these questions is given by M. Dine,
hep-th/9207045.
\bibitem{CLLSW} E. J. Copeland, A. R. Liddle, D. H. Lyth, E. D. Stewart
        and D. Wands, Phys. Rev. D {\bf 49}, 6410  (1994).
\bibitem{dinefisch} M. Dine, W. Fischler and D. Nemeschansky,
        Phys. Lett. {\bf 136B}, 169 (1984).
\bibitem{coughlan} G. D. Coughlan, R. Holman, P. Ramond and G. G. Ross,
        Phys. Lett. {\bf 140B}, 44 (1984).
\bibitem{ewansgrav} E. D. Stewart, Phys. Rev. D {\bf 51}, 6847 (1995).
\bibitem{casas} 
J.A. Casas and C. Munoz, Phys. Lett. {\bf B216}, 37 (1989);
J.A. Casas, J.M. Moreno, C. Munoz and M. Quiros, Nucl. Phys. {\bf B328}, 
272 (1989).
\bibitem{bindvali} P. Binetruy and G. Dvali,  
Phys. Lett. {\bf B388}, 241 (1996).
\bibitem{halyo} E. Halyo, Phys. Lett. {\bf B387}, 43 (1996).
\bibitem{j2} R. Jeannerot, hep-ph/9706391. 
\bibitem{dvaliriotto} G. Dvali and A. Riotto, hep-ph/9706408.
\bibitem{casas2} J. A. Casas and G. B. Gelmini, 
hep-ph/9706439.
\bibitem{lr} D. H. Lyth and A. Riotto, 
hep-ph/9707273.
\bibitem{steve} M. Bastero-Gil and S. F. King, 
hep-ph/9709502.
\bibitem{dss}  G. Dvali, Q. Shafi  and R. Schaefer,
Phys. Rev. Lett. {\bf 73}, 1886 (1994).
\bibitem{izawa} K.-I. Izawa and T. Yanagida,
hep-ph/9608359.
\bibitem{linderiotto} A. D. Linde and A. Riotto, hep-ph/9703209, 
to be published in Phys. Rev. {\bf D}.  
\bibitem{ddr} S. Dimopoulos, G. 
Dvali and R. Rattazzi, hep-ph/9705348.
\bibitem{natural} K. Freese, J. Frieman and A. V.  Olinto
Phys. Rev. Lett. {\bf 65}, 3233 (1990);
F. C. Adams et al.
1993, Phys. Rev. D {\bf 47}, 426 (1993);
L. Knox and A. Olinto, Phys. Rev. D {\bf 48}, 946
 (1993); W. H. Kinney and K. T. Mahanthappa,
Phys. Rev. D {\bf 52}, 5529 (1995);
W. H. Kinney and K. T. Mahanthappa,
 Phys. Rev. D {\bf 53}, 5455 (1996);
J. Garcia-Bellido, A. Linde and D. Wands, astro-ph/9605094.
\bibitem{gmo}  M. K. Gaillard, H. Murayama and K. A.
Olive, Phys. Lett. {\bf B355}, 71 (1995).
\bibitem{glm} M. K. Gaillard, H. Murayama and D. H. Lyth,
in preparation.
\bibitem{mtheory} E. Witten, Nucl. Phys. {\bf B471}, 135 (1996). 
\bibitem{banksdine} This observation has been made by T. Banks and M. Dine,
hep-th/9609046.
\bibitem{ewanloop2} E. D. Stewart, hep-ph/9703232.
\end{thebibliography}
\end{document}